\documentclass[
aps,
prl,
reprint,
superscriptaddress,
amsfonts,
amssymb,
amsmath,
showpacs,
eprint
]{revtex4-1}

\usepackage{graphicx}
\usepackage{dcolumn}
\usepackage{bm}
\usepackage[normalem]{ulem}
\usepackage{framed}
\usepackage{hyperref}
\usepackage{color}
\usepackage{braket}
\usepackage{textcomp}
\usepackage{nicefrac}
\usepackage{sidecap}
\usepackage{float}

\def\r1{\textbf{r}}

\usepackage{makeidx}
\makeindex

\begin{document}

\abovedisplayskip=8pt
\abovedisplayshortskip=8pt
\belowdisplayskip=6pt
\belowdisplayshortskip=6pt


\title{Single-photon interaction between two quantum dots located in different cavities of a weakly coupled double microdisk structure 
}
\author{S.~Seyfferle}
\altaffiliation{S. Seyfferle and F. Hargart contributed equally to this work. }
 
\affiliation{Institut f\"ur Halbleiteroptik und Funktionelle Grenzfl\"achen, Research Center SCoPE and IQ$^{\text{ST}}$, Universit\"at Stuttgart, Allmandring 3, 70569 Stuttgart, Germany}

\author{F.~Hargart}
\altaffiliation{S. Seyfferle and F. Hargart contributed equally to this work. }
\affiliation{Institut f\"ur Halbleiteroptik und Funktionelle Grenzfl\"achen, Research Center SCoPE and IQ$^{\text{ST}}$, Universit\"at Stuttgart, Allmandring 3, 70569 Stuttgart, Germany}
\thanks{S. Seyfferle and F. Hargart contributed equally to this work.}


\author{M.~Jetter}
\affiliation{Institut f\"ur Halbleiteroptik und Funktionelle Grenzfl\"achen, Research Center SCoPE and IQ$^{\text{ST}}$, Universit\"at Stuttgart, Allmandring 3, 70569 Stuttgart, Germany}

\author{E.~Hu}
\affiliation{School of Engineering and Applied Sciences, Harvard University, 29 Oxford Street, Cambridge, MA 02138, USA}

\author{P.~Michler}
\affiliation{Institut f\"ur Halbleiteroptik und Funktionelle Grenzfl\"achen, Research Center SCoPE and IQ$^{\text{ST}}$, Universit\"at Stuttgart, Allmandring 3, 70569 Stuttgart, Germany}

\date{\today}

\begin{abstract}
We report on the radiative interaction of two single quantum dots (QDs) each in a separate InP/GaInP-based microdisk cavity via resonant whispering gallery modes. The investigations are based on ab initio coupled disk modes.
We apply optical spectroscopy involving a $4f$-setup, as well as mode-selective real space imaging and photoluminescence mapping to discern single QDs coupled to a resonant microdisk mode. Excitation of one disk of the double cavity structure and detecting photoluminescene from the other yields proof of single photon emission of a QD excited by incoherent energy transfer from one disk to the other via a mode in the weak coupling regime. Finally, we present evidence of photons emitted by a QD in one disk that are transferred to the other disk by a resonant mode and are subsequently resonantly scattered by another QD. 
\end{abstract}


\maketitle


The scientific development towards quantum technologies  based on semiconductor solid-state devices has seen much progress in recent years. In particular, semiconductor quantum dots (QDs) put themselves forward for the implementation as qubits \cite{Loss.DiVincenzo_1998, doi:10.1063/1.1416855}. The coherent control of the interaction of two QDs in coupled quantum systems is a key element and promises, e.g., the implementation of parallel qubit operation for quantum information processing. To this end, Imamo\u{g}lu et al. proposed \cite{Imamoglu.Awschalom.ea_1999} to utilize two spatially distant electron spins of QD excitons inside a microcavity coupled by a single cavity field to implement CNOT-operations.

The coupling of two QDs in one cavity has been realized in micropillar \cite{Reitzenstein.Loeffler.ea_2006} and photonic crystal cavities \cite{Laucht.Villas-Boas.ea_2010,Kim.Sridharan.ea_2011}. However, accomplishing selective tunability and individual addressability of each QD, which is essential for the manipulation of individual qubits in future applications, is technically challenging for closely spaced QDs.
Consequently, the idea is to exploit the long-range interaction between QDs in coupled microcavity systems, i.e., photonic molecules (PMs) \cite{Bayer.Gutbrod.ea_1998, Senellart, PhysRevB.86.045315, Lin:10} where the energy transfer is mediated via a resonant cavity mode.
It has been shown recently that the excitation of a two level system with quantum light instead of classical light could improve the quality of subsequently emitted single photons \cite{Carreno.SanchezMunoz.ea_2016}. Thus, classical light could be used to excite a first QD which in turn excites a second QD with a stream of single photons. Using PMs to obtain an efficient coupling between the QDs presents an integrable and compact semiconductor solution.
In addition, these PMs have also been theoretically investigated for entanglement of a pair of QDs \cite{Vasco.Guimaraes.ea_2014,Vasco.Gerace.ea_2016}, they enable the unconventional photon blockade \cite{Liew.Savona_2010,Bamba.Imamoglu.ea_2011}, and serve as a starting point for the realization of driven-dissipative multi-cavity systems \cite{Gerace.Tureci.ea_2009} and strong photon-photon correlations \cite{Hartmann.Brandao.ea_2006,Greentree.Tahan.ea_2006,Rodriguez.Amo.ea_2016}.

An attractive type of cavity systems are whispering gallery mode (WGM) supporting microdisks (see exemplary SEM-picture in Fig. \ref{Fig1} a)). The WGMs propagate along the inner edge of the disk and couple to an adjacent microdisk cavity via the exponentially decaying evanescent mode field outside of the disk slab. The concentration of the electromagnetic mode field along the rim of the disks grants an effective photonic coupling to the QDs. Furthermore, the micrometer dimensions of the cavity allow for convenient optical addressability of each microdisk.

In this letter, we present resonant scattering of single photons by a QD, which have been emitted by another QD in the other disk of the double GaInP-based microdisk structure containing an embedded active layer of InP QDs. The radiative excitation transfer is mediated by a weakly coupled cavity mode. The WGM wavelength is very sensitive to the disk diameter leading to a mode energy mismatch between the disks because of imperfections in the fabrication process. These spectral mode differences can be compensated by local tuning techniques, e.g., laser heating \cite{Benyoucef.Kiravittaya.ea_2008,Witzany.Liu.ea_2013}, electro-thermal heating \cite{Faraon.Vuckovic_2009}, photo-reactive materials \cite{Faraon.Englund.ea_2008}, which are all based on the local control of the refractive index, or photo-electrochemical etching \cite{Gil-Santos.Baker.ea_2017} to change the cavity dimension on the nanometer scale.
However, most tuning techniques are experimentally very challenging or prevent single QD experiments, e.g., by high excitation powers necessary for sufficient cavity mode tuning.
For that reason we follow a different approach by discerning modes which \textit{ab initio} display evidence of coupling, in order to be independent of tuning mechanisms.  
We identify possibly coupled microdisk modes by means of micro-photoluminescene (\textmu -PL) spectroscopy scans in combination with real space imaging, verify single-photon emission from QDs excited by energy transfer via mode coupling and present indications of on-chip resonant scattering.


\section{Sample and setup}
The sample exclusively studied in the work at hand was grown by metal-organic vapor-phase epitaxy (MOVPE). 
The microdisk post is made of Al$_{0.5}$Ga$_{0.5}$As:Si. The disk features layers of 
GaInP and AlGaInP symmetrically arranged around the active layer of InP QDs that have a density of $1.5 \cdot 10^{10}\,\text{cm}^{-2}$.
The actual disk structure is finally formed by a combination of e-beam lithography, dry etching and wet undercut etching. Fabrication and processing steps are detailed in \cite{Witzany.Liu.ea_2013}.
The disk dimers have a nominal interdisk minimum edge separation of $50\,\text{nm}$ while each disk possesses a diameter of $5\,\text{$\mu$m}$. Typical $Q$-factors of the sample range in the order of 12000. High $Q$-factors are desirable to grant an efficient QD-mode-coupling. In the so-called strong coupling regime a coherent and reversible exchange of photons between QD and cavity mode is given which outweighs the cavity losses \cite{Forchel, PhysRevLett.95.067401}. 
However, the following investigations are based on the weak coupling of both, QD to a mode as well as the coupling of two modes in different cavities.
An estimate of the upper limit of the Purcell factor for a QD coupled to a disk mode based on the above mentioned Q-factor and an exemplarily calculated mode volume of $V_{\text{mode}} \approx 17 (\lambda / n)^3$ can be given by $F_{\text{P}} \approx 54$. Furthermore, the coupling strength can be estimated to $g \approx 62\,$\textmu eV using \cite{Srinivasan:06}
\begin{equation*}
\frac{g}{2 \pi}= \frac{1}{2 \tau} \sqrt{\frac{3c\lambda ^2 \tau}{2 \pi n^3 V_{\text{mode}}}}
\end{equation*}
with a typical radiative lifetime of InP QDs of $\tau =350\,$ps \cite{PhysRevB.75.195302} and a refractive index of the GaInP-resonator of $n\approx 3.6$. Comparing $g$ to the loss rate $\kappa = \omega /Q \approx 153\,$\textmu eV confirms that the system is in the weak coupling regime. In this regime medium Q-factors benefit the mode-mode-coupling \cite{Benyoucef.Kiravittaya.ea_2008,Witzany.Liu.ea_2013, Lin:10} between two different microdisks since a larger fraction of photons escapes the cavity to interact with a resonant mode in the other. Therefore, also in the weak coupling regime resonant energy transfer via cavity modes and dot-to-dot interaction is possible.
\begin{figure}
        \centering
                \includegraphics[width=0.48\textwidth]{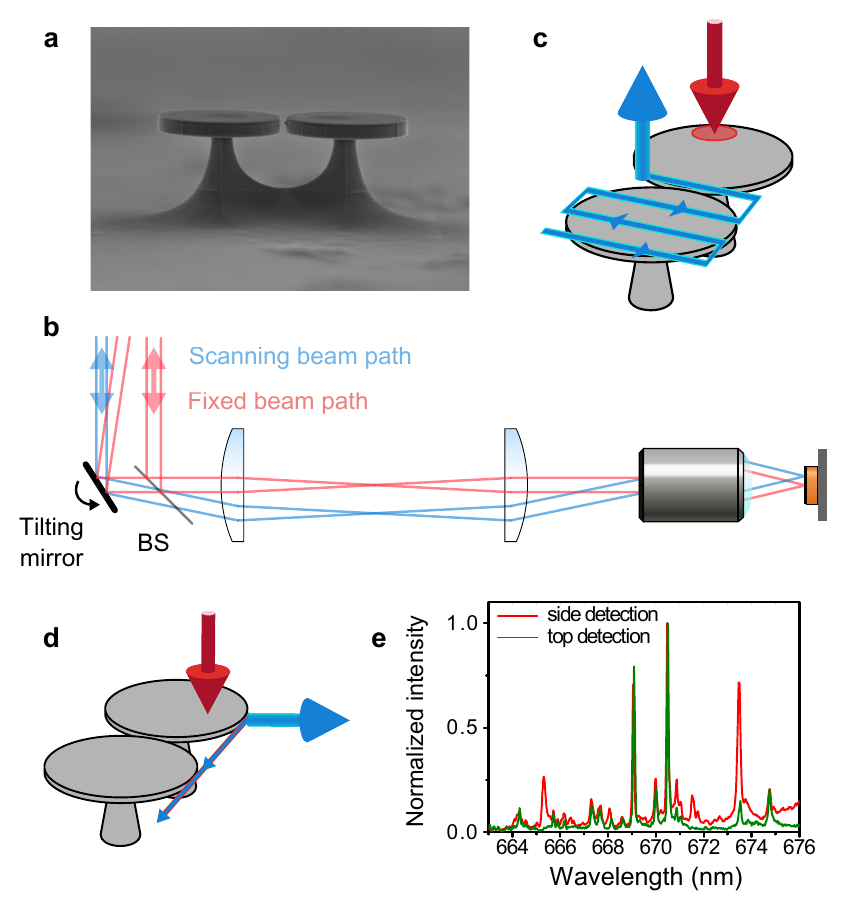}
        \caption{\label{Fig1}
        (Color online) a) Exemplarily SEM image of a microdisk dimer with a nominal interdisk spacing of 50\,nm.
        b) $4f$-setup for PL scans and maps. The lens configuration in combination with a piezo driven tiltable mirror enables to scan detection and excitation independently from each other over the disk pair.
				c) Schematic of the PL map scanning procedure, here, for fixed excitation (red) and scanning detection (blue). The setup also allows for other configurations, i.e., scanning both the excitation and detection, or fixed detection and scanning excitation.
				d) Schematic of the PL side scanning procedure. The cryostat geometry makes possible to optically access the sample also in-plane.
				e) Comparison of spectra taken from the disk side (red solid line) and disk top (green solid line). 
        }
\end{figure}

For the spectral characterization of the microdisks via \textmu -PL spectroscopy, the sample is placed inside a helium flow cryostat on a motorized stage at a nominal temperature of $4\,\text{K}$. 
The above band excitation laser ($532\,\text{nm}$) is focused onto the outer disk rim through a $100$x microscope objective ($\text{NA}=0.7$). The focused spot diameter is estimated to $<1.5\,\text{\textmu m}$, thus largely excluding the possibility of involuntary excitation of QDs in the other disk, assuming a Gaussian laser intensity profile.
The PL of the disks is collected by the same objective and passes through a $4f$ lens system of which a sketch is shown in Fig.~\ref{Fig1} b). This top detection scheme mainly collects stray light of the mode emission since the modes emit predominantly radially and in-plane. The $4f$ setup involving two convex lenses ($f=20\,\text{cm}$) and a piezo driven tiltable mirror allows for the spatial disconnection of excitation and detection beam paths in a confocal microscopy setup. 
This configuration makes possible either local invariant excitation of one disk of the dimer while simultaneously and independently scanning the detection across the complete disk pair by tilting the piezo mirror, or alternatively, moving the excitation spot while using a fixed detection spot. 
In this way, the whole disk pair can be scanned by stepwise movement of the tiltable mirror and recording a spectrum for each mirror position, as is schematically drawn in Fig. \ref{Fig1} c). This enables to record one-dimensional scans as well as two-dimensional photoluminescence maps. The cryostat geometry also enables to optically access and scan the sample from the side (see schematic Fig. \ref{Fig1} d)) thereby detecting mainly the in-plane mode emission. 

For the purpose of real space CCD-imaging the PL-signal from the sample is collected by the same objective and focused via a collective lens onto a liquid nitrogen cooled CCD-Chip. Insertion of suitable bandpass filters enables to image only certain modes of interest.

A basic optical characterization of a disk pair is shown in Fig.~\ref{Fig1} e).
Spectra detected from the top of the disks (perpendicular to the sample surface, green solid line) and from the disk sides (in-plane, red solid line) are plotted for comparison. The disk has been excited from the top and the spectra have been obtained by directly switching between the two orthogonal detection configurations without varying the setup further. The spectra show several sharp and bright emission lines likely due to high-Q modes fed by QDs interspersed with weaker and broader lines probably caused by modes of lower Q-factor. Due to the predominant radial and in-plane emission characteristics of the WGM, it is expected to directly access mode emission from the side while only detecting mode stray light from the top. The two dominating lines at 669\,nm and 670.4\,nm appear almost equally bright in both detection schemes, whereas the peak at 673.5\,nm displays higher intensity detected from the side. Consequently stray light collection for this mode from the top is less efficient. 
Additionally, there are two lines at 665.3\,nm and 671.5\,nm which only appear in the side detection and can assumed to be disk modes whose straylight is not scattered into the top microscope objective. The overall similar appearance of the two spectra shows that straylight detection from the top is a convenient way for the optical investigations of the microdisk pairs especially with respect to imaging and mapping the mode profile from the top seen in the following.


\section{Imaging of coupled and uncoupled modes}
Taking one-dimensional \textmu -PL spectroscopy scans as a starting point enables to pre-select possibly coupled disk modes by the observation of emission lines that display a considerable intensity at the same wavelength in both disks.
\begin{figure}
        \centering
                \includegraphics[width=0.48\textwidth]{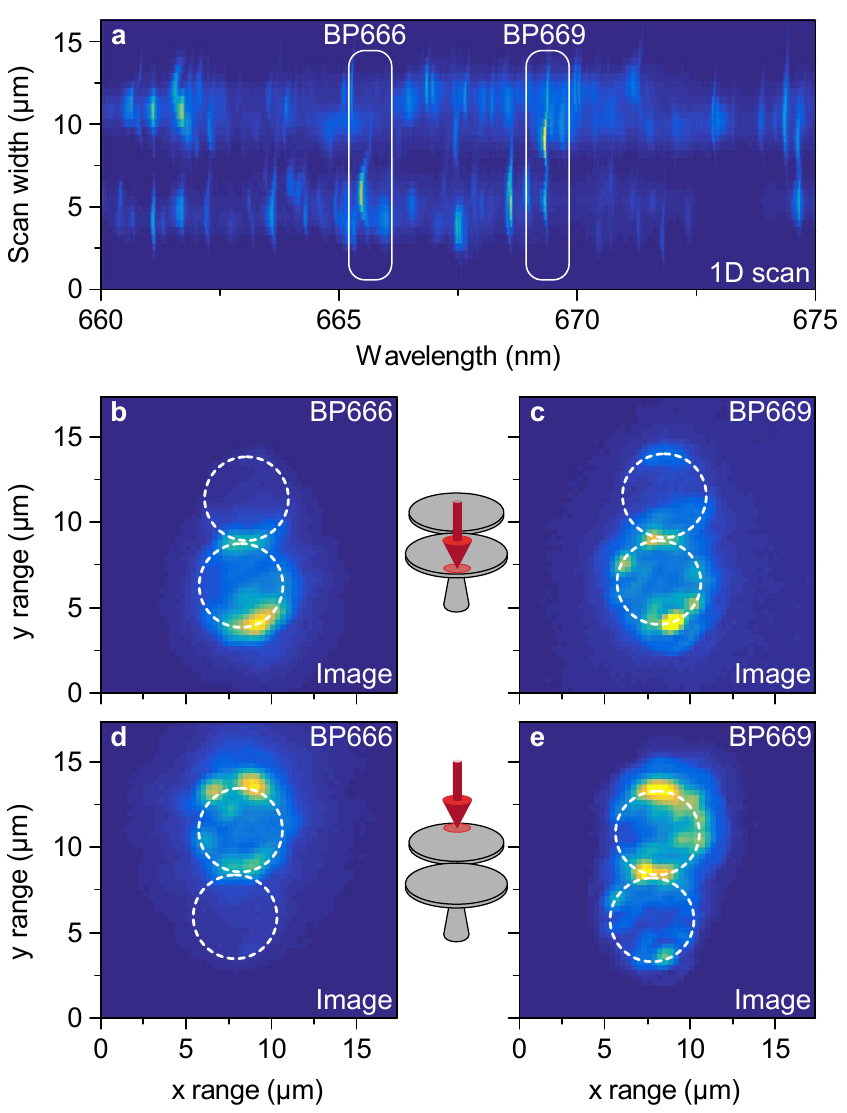}
        \caption{\label{Fig2}
        (Color online) a) Normalized color-scale plot of a one-dimensional scan displaying emission lines from both disks. Two emission lines signified in the plot are selected for further investigations: a presumably coupled mode at $\approx 669\,\text{nm}$ and an uncoupled mode at $\approx 666\,\text{nm}$. b) - e): Real space images of the disk pair evaluated with a bandpass filter at $669\,\text{nm}$ ($666\,\text{nm}$) for the (un)coupled mode. The lower disk is excited in b) and c) (see inset). Only in case of the supposedly coupled mode the non-excited disk contributes PL. In d) and e) the upper disk is excited. Again, the non-excited disk remains dark for the uncoupled mode. The red arrows in the insets show the spatial position of the excitation laser on the disk.
        }
\end{figure}

Fig.~\ref{Fig2} a) shows the scan of a dimer displaying a number of emission lines whose spatial distribution clearly outline the individual disk location. The scan was obtained by stepwise motion of the above band excitation laser with an excitation power of $60\,$nW over the length of the dimer while acquiring a spectrum at each laser position. 
 Some emission lines can be seen to be situated at the same wavelength in both disks, which can be taken as a first hint at resonant modes, which in turn is a precondition for radiative coupling. Such a candidate is evident at $669\,\text{nm}$. A counter-example can be found at $666\,\text{nm}$ where the intensity of the emission line is predominantly restricted to the disk in the bottom scan part.
Both, the presumably coupled ($669\,\text{nm}$) and uncoupled mode ($666\,\text{nm}$) have been selected for further investigations by real space imaging (cf. Fig.~\ref{Fig2} b) - e)). The respective modes have been selected with an appropriate bandpass filter (transmission window $\pm 0.5\,\text{nm}$). 
The images in Fig.~\ref{Fig2} b) and c) have been recorded with the excitation laser fixed on the far edge of the bottom disk (see schematic insets in Fig.~\ref{Fig2}). In case of the presumably uncoupled mode at $666\,\text{nm}$ the lower disk reveals distinct centers of emission along the expected circular mode profile (Fig.~\ref{Fig2} b)), whereas the upper disk does not contribute substantial PL, which supports the assumption, that this mode has no resonant counterpart in the other disk. 
However, a ring of PL emission outlining the disk shape can be perceived in Fig.~\ref{Fig2} c) in case of the presumably coupled mode at $669\,\text{nm}$. 
This indicates radiative excitation transfer from the excited disk to the non-excited.

A cross-check has been performed by placing the laser spot on the other disk (Fig.~\ref{Fig2} d) and e)). Again, the non-excited disk remains dark for the presumably uncoupled mode at $666\,\text{nm}$ (Fig.~\ref{Fig2} d)), i.e. no excitation transfer from one disk to the other takes place. Note, that in both disks QDs are present that emit around a wavelength of $666\,\text{nm}$, but only as a consequence of direct optical excitation by the laser. This observation should exclude other excitation transfer processes, e.g., via phonons or scattering processes. 
The same procedure has been carried out for the presumably coupled mode (Fig.~\ref{Fig2} e)). The non-excited lower disk is clearly visible by a ring of emission along the disk edge where the mode is expected to propagate, as well as bright circular centers of high intensity which indicate QDs excited by transferred light. 
Another feature that appears in the images of Fig.~\ref{Fig2} c) and e) is a striking bright luminescence at the interdisk gap, which can be attributed to electric field enhancement at the contact point and can be seen as an additional indication of resonance of the modes in the two disks. This phenomenon is comparable to plasmonic \textit{hot spots} \cite{Sigalas.Fattal.ea_2007}.
 
The spots of high intensity seen in the real space images in Fig.~\ref{Fig2} can be identified as (single) QDs emitting preferentially perpendicular to the disk surface. This way individual QDs can be conveniently selected for single dot experiments.


\section{Single QD coupling to resonant disk modes}
\begin{figure}
        \centering
                \includegraphics[width=0.48\textwidth]{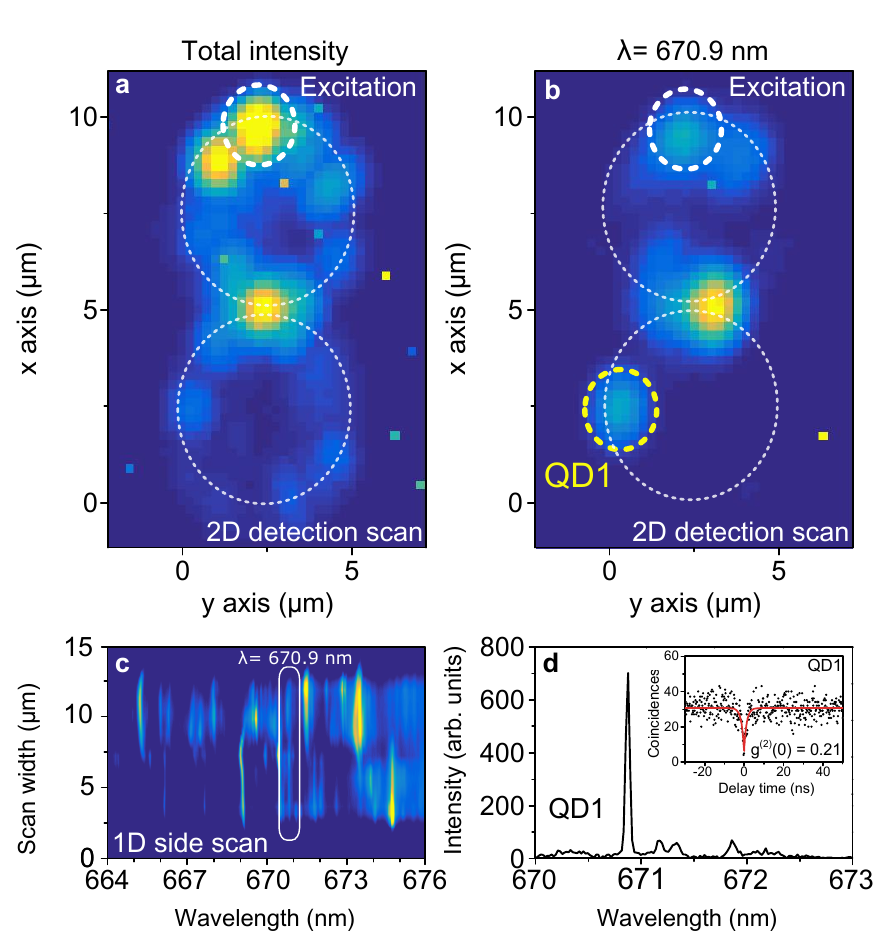}
        \caption{\label{Fig3}
        (Color online) a) Color-scale plot of a two-dimensional \textmu PL-scan applying an excitation power of $94.5\,$nW with the excitation spot fixed on the upper disk of the dimer, indicated by the dotted white circle. The ensemble map shows the complete integrated intensity. A distinct mode profile can be seen from both the excited and non-excited disk as well as field enhancement at the inter-disk gap. The dashed circles serve as guides to the eye and indicate the disk edges. b) Spectrally post-selected intensity map at $670.9\,$nm of the same measurement as in a). The non-excited lower disk shows a bright spot indicating a single QD excited by energy transfer via a resonant mode. c) A one-dimensional \textmu PL-scan from the side of the disks reveals a resonant mode at $670.9\,$nm responsible for the energy transfer that excites \textit{QD1}. d) The spectrum taken at the spatial position of \textit{QD1} is depicted a sharp emission line. Auto-correlation measurement on this line. Note: The excitation laser is still focused on the upper disk. The anti-bunching dip indicates a single QD excited via a coupled microdisk mode.
        }
\end{figure}

For more detailed investigations of possible single dot coupling in the microdisk systems two-dimensional PL maps have been recorded applying the $4f$-setup described above. These maps contain the complete spectral information at each spatial coordinate on the disks. A resulting ensemble map of such a $x$-$y$-scan with a fixed excitation spot on the upper disk is pictured in Fig.~\ref{Fig3} a).
Several features catch the eye: The bright spot marked with the white dotted circle indicates the position of the excitation laser. The non-excited disk is represented in the map by the WGM profile outlining the disk geometry and indicating coupling. Again, in the gap between the disks the field is strongly enhanced. A spectrally post-processed version of this measurement is shown in Fig.~\ref{Fig3} b) which represents the same disk pair filtered at a wavelength of $670.9\pm 0.1\,\text{nm}$. A high intensity spot (referred to as \textit{QD1}) can be seen on the non-excited disk leading to the assumption of a QD that is excited by radiative transfer of excitation from the excited disk to the non-excited mediated by a resonant disk mode. The \textmu PL-map in Fig.~\ref{Fig3} b) does not show distinct intensity along the disk outlines which could be assigned to a resonant mode at $670.9\,\text{nm}$. Therefore, a one-dimensional \textmu PL-scan along the disk sides has been carried out. The result in Fig.~\ref{Fig3} c) reveals an emission line at $670.9\,\text{nm}$ which is present in both disks highlighted by the white box. Consequently, the resonant mode  responsible for the energy transfer that excites \textit{QD1} can only be detected from the disk side due to the predominant radial in-plane emission characteristics and in this case low detection of mode stray light.  
\begin{figure}
        \centering
                \includegraphics[width=0.48\textwidth]{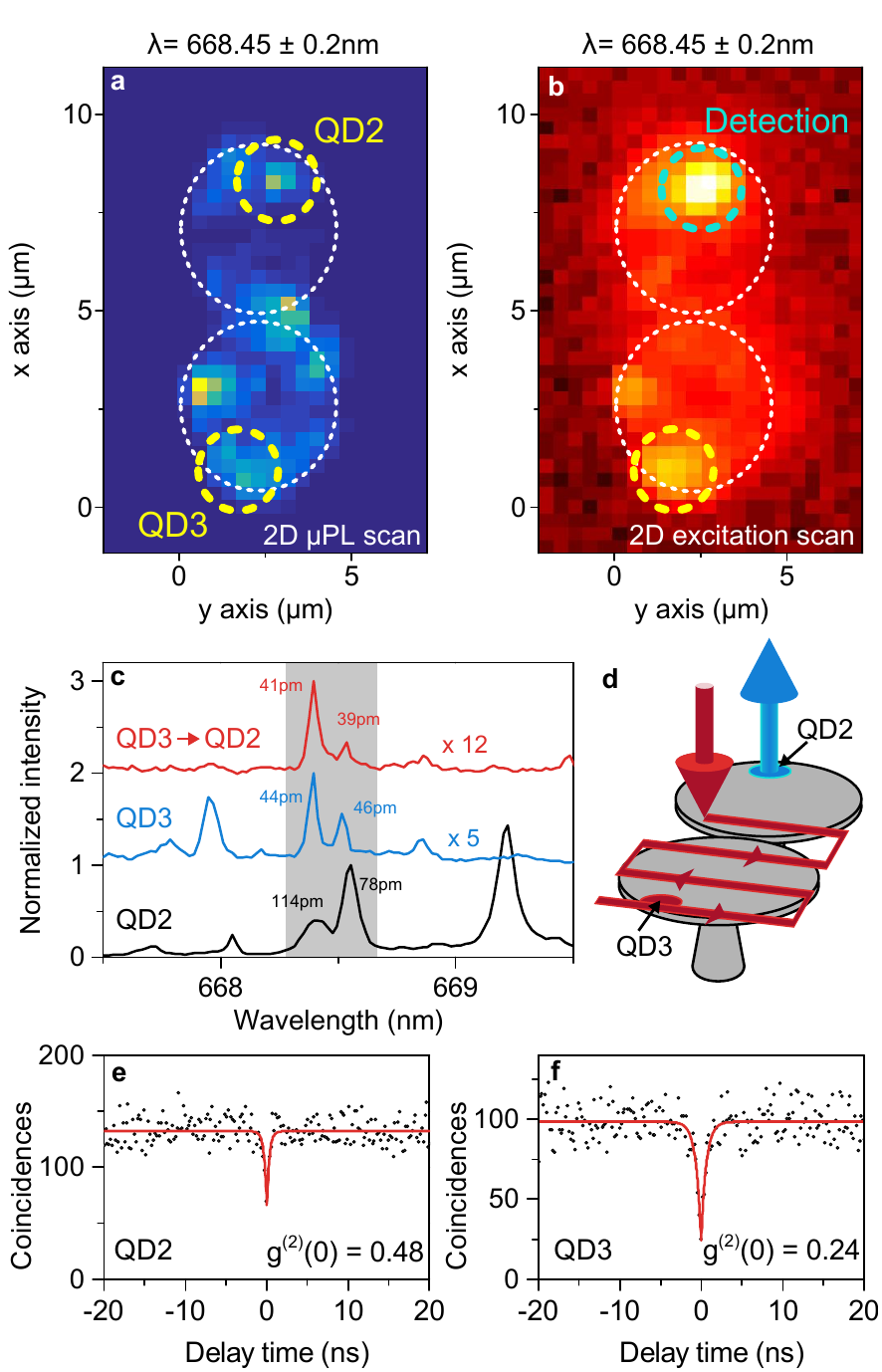}
        \caption{\label{Fig4}
        (Color online) a) Normalized linear color-scale plot of a two-dimensional \textmu PL scan with both excitation and detection moving across the microdisk dimer. The post-selected intensity for a spectral region of $668.4\pm0.2\,$nm is displayed. Two QDs with almost the same emission wavelength are found in the different disks, labeled with \textit{QD2} and \textit{QD3} (see subfigure c) for their spectra).
				b) Normalized logarithmic color-scale plot of a 2D excitation scan, i.e., the detection is spatially fixed on \textit{QD2} while the excitation laser is scanned across the cavity structure (see subfigure d). We plot the intensity for the same spectral region and indicate a strong intensity increase when scanning over \textit{QD3} in the lower disk.
				c) Comparison of the two ordinary \textmu PL spectra for \textit{QD2} (upper disk) and \textit{QD3} (lower disk) with the spectrum recorded in the upper disk when exciting \textit{QD3} in the lower disk. The highlighted area in grey outlines the filtered spectral region in the \textmu PL maps of subfigures a) and b). The inset numbers denote the FWHM values of each peak.  
				e) \& f) Auto-correlation measurements indicating antibunching on \textit{QD2} and \textit{QD3}, respectively. Both measurements have been taken at the spectral peak on the right hand side of c)
        }
\end{figure}
 Without varying the position of the excitation, the emission spectrum is obtained at the location of the spot in Fig.~\ref{Fig3} b) (\textit{QD1}) and displayed in Fig.~\ref{Fig3} d). The high intensity peak represents the emission spot seen in the map of Fig.~\ref{Fig3} b). An auto-correlation measurement of this emission line yielded the resulting coincidence histogram depicted in Fig.~\ref{Fig3} d). Clear antibunching behavior can be observed from the measurements with a $g^{(2)}(0)$ value of $0.21$, indicating a single emitter. This is evidence of single dot excitation in the non-excited disk mediated by mode coupling. It is necessary to point out, that  no exclusive dot-to-dot excitation takes place in this measurement, since the spectrum taken at the excitation spot (not shown) reveals considerable background emission and no single QD line due to a relatively high excitation power. Such a dot-to-dot interaction will be shown in the next paragraph.


\section{On-chip resonant scattering of two QDs in different microdisks}
Fig.~\ref{Fig4} a) shows the colour plot of a \textmu -PL map filtered post-selectively at a wavelength of $668.45 \pm 0.2\,\text{nm}$. Excitation (power: $80\,$nW) and detection have been scanned simultaneously above the disk surface, acquiring a spectrum at every excitation spot. Several centers of bright emission intensity can be seen in the map, two of them are highlighted by the yellow dotted circles in Fig.~\ref{Fig4} a). These can be again identified as single QDs by auto-correlation measurements (see Fig.~\ref{Fig4} e) \& f)) and serve as objects of the following study. Their normalized spectra are displayed in Fig.~\ref{Fig4} c), labeled as \textit{QD2} and \textit{QD3} according to Fig.~\ref{Fig4} a). Each spectrum reveals two emission lines in the highlighted area (spectral filter in Fig.~\ref{Fig4} a) and b)) with the same emission wavelength, a precondition for radiative interaction. The spectral position and the linewidths can be extracted from the data using Gaussian fit functions. The lines of \textit{QD2} have a wavelength of $668.41\,\text{nm}$ and $668.55\,\text{nm}$ and linewidths of $114\,\text{pm}$ and $78\,\text{pm}$, respectively. The two lines of \textit{QD3} are located at $668.39\,\text{nm}$ and $668.52\,\text{nm}$ with linewidths of $44\,\text{pm}$ and $46\,\text{pm}$, respectively. The particular single lines display an energy splitting of $\approx 400\,$\textmu eV and arise most probably either from different charge configurations in the QDs \cite{Reischle:08} or from a distinct fine-structure splitting of the QD exciton state \citep{PhysRevB.59.R5300}.
Fig.~\ref{Fig4} b) displays the logarithmically plotted \textmu -PL map of the same disk pair, only in this instance the excitation has been scanned across the disk while the detection remained fixed to the spatial position of \textit{QD2}, as schematically depicted in Fig.~\ref{Fig4} d). The plot features again the two previously observed emission spots if compared to the \textmu -PL map in Fig.~\ref{Fig4} a).
The remarkable findings of this observation is, that as soon as the laser excites \textit{QD3}, the collected emission at \textit{QD2} strongly increases. The corresponding spectrum is also shown in Fig.~\ref{Fig4} c) as \textit{QD3} $\rightarrow$ \textit{QD2}. Note, only the exact same lines from \textit{QD3} are detected which have a resonant counterpart at \textit{QD2}. All other lines, e.g., at roughly 668\,nm and 669.15\,nm are not observable. Most interestingly, the measured line shapes detected at \textit{QD2} when \textit{QD3} is excited ($41\,\text{pm}$ and $39\,\text{pm}$) correspond to the directly excited \textit{QD3}, while they are dissimilar to the FWHM of \textit{QD2} excited on-site.
This suggests that emitted photons from \textit{QD3} are transferred via a cavity mode to QD2 and are subsequently scattered by this two-level system or in other words on-chip resonant scattering of quantum light from one QD by another. 
Differences in the intensity might occur due to different coupling strengths and scattering cross sections. 
The coupling efficiency can be roughly estimated from the count rates. This gives an indication of the fraction of the photons that are emitted by \textit{QD3} and scattered at \textit{QD2}. As indicated in Fig.~\ref{Fig4} c) the difference in count rates is a factor of 2.4 and thus the coupling efficiency accounts to $\approx 40 \%$. This considerable value is probably due to the high Q-factor, which enhances the capture and storage of photons into the cavity modes and thus increases the likelihood of transfer into the other disk and subsequent excitation of \textit{QD2}.


\section{Conclusion and outlook}

In conclusion, we have demonstrated the identification of resonant cavity modes by \textmu PL scans and real space imaging. Using such modes, the excitation of single QDs located in another disk is possible and was proved by auto-correlation measurements.
Finally, we presented indications of photons emitted from one QD which are transferred to the other disk via a resonant mode where they are resonantly scattered by another QD.

For future applications and a more detailed investigation of our basic results, we have to reinstate and further develop advanced tuning mechanisms to control the QD emission wavelength and, thus, the scattering mechanism.

\begin{acknowledgments}
The authors gratefully ackowledge financial support by the Deutsche Forschungsgemeinschaft via the grant SFB/TRR21.
We thank T.-L. Liu for processing the high quality sample.
\end{acknowledgments}

\bibliography{bibfile}
\bibliographystyle{unsrt}

\printindex

\end{document}